\documentclass[3p,times,twocolumn]{elsarticle}

\usepackage{ecrc}

\volume{00}

\firstpage{1}

\journalname{Nuclear and Particle Physics Proceedings}

\runauth{Christian Bierlich}

\jid{nppp}

\jnltitlelogo{Nuclear and Particle Physics Proceedings}

\usepackage{amssymb}
\usepackage[figuresright]{rotating}

\usepackage{axodraw}
\usepackage{amsmath}
\usepackage{epsfig}
\usepackage{xspace}
\usepackage{graphics}
\usepackage[utf8]{inputenc}
\usepackage{placeins}
\usepackage{ifthen}

\newcommand{\dipsy}{\protect\scalebox{0.8}{DIPSY}\xspace}
\newcommand{\ariadne}{\protect\scalebox{0.8}{Ariadne}\xspace}
\newcommand{\pytppp}{P\protect\scalebox{0.8}{YTHIA}8\xspace}

\def\pmb#1{{\mbox{\boldmath$#1$}}}
\def\eg{\emph{e.g.}}
\def\ie{\emph{i.e.}}

\def\etal{\emph{et al}}
\begin{document}

\begin{frontmatter}



\dochead{}

\title{Multiparton interactions: From pp to pA\tnoteref{label1}}
\tnotetext[label1]{Based on work done with G{\"o}sta Gustafson and Leif L{\"o}nnblad. This  work  was  supported  in  part  by  the  MCnetITN  FP7  Marie  Curie  Initial  Training Network,  contract PITN-GA-2012-315877,  the Swedish Research Council (contracts 621-2012-2283 and 621-2013-4287). Preprint number: LU-TP 16-58, MCNET-16-40.}

\author{Christian Bierlich}

\address{Dept. of Astronomy and Theoretical Physics, S{\"o}lvegatan 14A S-223 62 Lund, Sweden}
\ead{Christian.Bierlich@thep.lu.se}
\begin{abstract}
In the process of understanding nuclear collisions, reliable extrapolations from pp collisions, based on Glauber models, are highly desirable, though seldomly accurate. We review the inclusion of diffractive excitations and argue that they provide an important contribution to centrality observables in pA collisions. We present a method for distinguishing between diffractively and non-diffractively wounded nucleons, and a proof-of-principle for an extrapolation of multiparton interaction models built on this.
\end{abstract}


\end{frontmatter}

% \linenumbers

\section{Introduction}
\label{sec:intro}
An important step towards fully understanding signals of QGP formation and collectivity in heavy ion collisions, is providing realistic extrapolations of the dynamics of pp collisions. Collisions of protons with nuclei is an important stepping stone, as the full nuclear geometry is already involved here, but the situation remains somewhat simpler than a full AA collision, as the number of sub-collisions is equal to the number of wounded nucleons in the target. 

In ref.~\cite{Bierlich:2016smv} we argued that the approximations normally used when extrapolating pp dynamics to pA collisions are too crude. We will present inclusion of fluctuations to the Glauber formalism, giving rise to a "wounded" cross section with contributions from diffractive excitations. We compare inclusion of fluctuations calculated in the \dipsy model with those from the Glauber--Gribov model, and use them to calculate distributions of wounded nucleons at LHC energies. Finally we present a simple model based on these principles, which allows for extrapolation of multiparton interaction models to pA, and comparisons to data. 

\section{Including fluctuations in pA collisions}
\subsection{Fluctuations in proton--proton}
The \dipsy \cite{Flensburg:2011kk} model is a dynamic initial state model, built on the Mueller dipole model \cite{Mueller:1994jq}. The model includes dynamics up to LL-BFKL, plus additional corrections from satuation and momentum conservation\footnote{The \dipsy initial state model is implemented in a full event generator, with final state radiation from the \ariadne shower \cite{Lonnblad:1992tz}. Hadronization is carried out by \pytppp \cite{Sjostrand:2014zea}, with added coherence effects from rope hadronization \cite{Bierlich:2014xba}. For more information, visit \texttt{http://home.thep.lu.se/DIPSY}.}.
The initial state is built up through evolution from an initial proton consisting of three valence dipoles. The evolution is in impact-parameter space and rapidity, with the dipole splitting probability per unit rapidity:
\begin{equation}
  \label{eq:dipoleradiation}
  \frac{d\mathcal{P}_g}{dY}=\frac{N_c \alpha_s}{2\pi^2}d^2\pmb{x}_g
\frac{(\pmb{x}_1-\pmb{x}_2)^2}{(\pmb{x}_1-\pmb{x}_g)^2 (\pmb{x}_g-\pmb{x}_2)^2}.
\end{equation}
An emission produces two new dipoles, $(\pmb{x}_1,\pmb{x}_g)$ and
$(\pmb{x}_g,\pmb{x}_2)$. The interaction probability between two dipoles, one from the left-moving cascade and one from the right-moving cascade:
\begin{equation}
  \label{eq:dipoleinteraction}
P=\frac{\alpha_s^2}{4}\left[\ln\left(
\frac{(\pmb{x}_1-\pmb{x}_3)^2(\pmb{x}_2-\pmb{x}_4)^2}
{(\pmb{x}_1-\pmb{x}_4)^2(\pmb{x}_2-\pmb{x}_3)^2}\right)\right]^2.
\end{equation}
Using the optical theorem, cross section can now be calculated. With convenient normalization, the optical theorem in impact parameter space reads:
\begin{equation}
	\Im{(A_{el})} = \frac{1}{2}(|A_{el}|^2 + P_{abs}).
\end{equation}
Where "abs" is short for "absorption", \ie~inelastic non-diffractive contributions. By inserting from equation (\ref{eq:dipoleinteraction}), we obtain the real elastic amplitude in impact parameter space, including all fluctuations in projectile and target:
\begin{equation}
	\label{eq:tofB}
	T(b) \equiv -iA_{el} =  1 - \exp\left(-\sum_{ij} f_{ij}\right).
\end{equation}
Since fluctuations are related to diffraction through the Good-Walker formalism, calculation of several semi-inclusive proton-proton cross sections is possible with equation (\ref{eq:tofB}). Here the absorptive, single diffractive and double diffractive:
\begin{align}
	\frac{d\sigma_{abs}}{d^2b} &= 2\left< T(b)\right> - \left< T(b)\right>^2,\\
	\frac{d\sigma_{SD,(p|t)}}{d^2b} &= \left<\left<T\right>^2_{(t|p)}\right>_{(p|t)} - \left<T\right>^2_{p,t},\\
	\frac{d\sigma_{DD}}{d^2b} &=\left<T^2\right>_{p,t} - \left<\left< T\right>^2_t\right>_p - \left<\left< T\right>^2_p\right>_t + \left<T\right>^2_{p,t}.
\end{align}
where subscripts $p$ and $t$ indicates averages over projectile and target respectively. The \dipsy formalism applies for protons and nuclei alike, and has recently been applied directly to pA collisions \cite{Gustafson:2015ada}. In this work \dipsy is only used to calculate fluctuations in the proton transverse structure, and a simpler model is used for extrapolation to pA (see section \ref{sec:extra}).

\subsection{Parametrization of cross section fluctuations}
When extrapolating from pp to pA collisions, it is important to keep in mind that many measurements will rely on a centrality measure, \eg~particle production in the forward (Pb going) direction. The relevant semi-inclusive cross section for an interaction between projectile and target will therefore have contributions from absorptive, single diffractive and double diffractive processes. We dub this the "wounded" cross section, inspired by the wounded nucleon model by Bia\l as \etal~\cite{Bialas:1976ed}:
\begin{figure}
	\centering
	\includegraphics[width=0.4\textwidth]{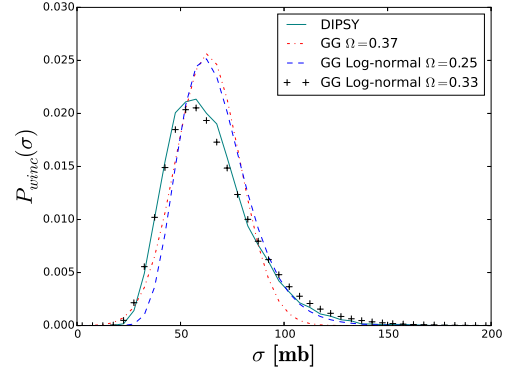}
	\caption{\label{fig:sigwounded}Fluctuations in $\sigma_w$ for \dipsy and three versions of GG fluctuations. GG fluctuations using a log-normal parametrization of $P_{tot}(\sigma)$ seems to be able to describe the \dipsy fluctuations best.}
\end{figure}
\begin{equation}
\frac{d\sigma_w}{d^2b} = \frac{d\sigma_{abs}}{d^2b} + \frac{d\sigma_{SD,t}}{d^2b} +  \frac{d\sigma_{DD}}{d^2b} =  2\left<T\right>_{p,t} - \left<\left<T\right>^2_t\right>_p.
\end{equation}
We can now compare the fluctuations in $\sigma_w$ produced by \dipsy with the often used parametrization, Glauber--Gribov Colour Fluctuations (GG) \cite{Alvioli:2013vk}. Here the fluctuations are parametrized with a distribution $P_{tot}(\sigma)$ such that:
\begin{align}
	\sigma_{tot} & = \int d\sigma  \sigma P_{tot}(\sigma) \\ 
		     & = \int d\sigma \rho \frac{\sigma^2}{\sigma + \sigma_0}\exp\left[-\frac{(\sigma/\sigma_0 -1)^2}{\Omega^2} \right], \label{eq:pstrik}
\end{align}
where the usual choice for $P_{tot}(\sigma)$ has been inserted in equation (\ref{eq:pstrik}). The parameters of the model can be fitted to semi-inclusive cross sections by assuming a functional form for $T$. Here we use a semi-transparent disk with:
\begin{equation}
	T(b,\sigma) = T_0 \Theta\left(\sqrt{\frac{\sigma}{2\pi T_0}} - b\right).
\end{equation}
The semi-inclusive cross sections are:
\begin{align}
	\sigma_{tot} &= \int d^2b \int d\sigma P_{tot}(\sigma)2T(b,\sigma) \\
	\sigma_{el} &= \int d^2b \left|\int d\sigma P_{tot}(\sigma) T(b,\sigma) \right|^2\\
	\sigma_{w} &= \int d^2b \int d\sigma P_{tot}(\sigma) \left[ 2T(b,\sigma) - T(b,\sigma)\right]
\end{align}
In figure \ref{fig:sigwounded} we show fluctuations in $\sigma_w$ at $\sqrt{s_{NN}} = 5.02$ TeV with \dipsy as well as GG, compared to GG with an modified $P_{tot}$ distribution -- a log-normal distribution -- which we find describes the \dipsy fluctuations better:
\begin{equation}
 P_{tot}(\sigma,b) = \frac{1}{\Omega \sqrt{2\pi}}\exp \left(-\frac{\log^2(\sigma/\sigma_0)}{2\Omega^2} \right).
\end{equation}
\begin{figure}
	\centering
	\includegraphics[width=0.4\textwidth]{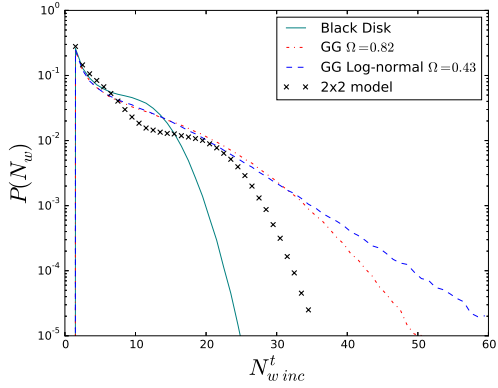}
	\caption{\label{fig:npart}Number of participants in collisions of pPb at 5.02 TeV using four different Glauber models.}
\end{figure}
\subsection{Extrapolation to pA}
\label{sec:extra}
In order to model pA collisions using the pp results from the previous section, we use a Woods-Saxon potential in the GLISSANDO parametrization \cite{Rybczynski:2013yba}. Colliding a proton with a lead nucleus at $\sqrt{s_{nn}} = 5.02$ TeV using the $\sigma_w$ as the relevant cross section is, however, not enough to proceed, as we want to distinguish absorptively wounded nucleons from diffractively excited ones. This is not possible in a GG parametrization, as it does not allow for separate fluctuations in projectile and target. This is accomodated in a crude way by allowing nucleons to fluctuate between two sizes with a fixed probablity. The real elastic amplitude for a projectile with radius $R_p$ to collide with target $R_t$ is $T(b) = \alpha\Theta(R_p + R_t - b)$, where $\alpha$ is an opacity. Adding a parameter ($c$) for the fluctuation probablity, the model has a total of four parameters ($r_1$, $r_2$, $c$ and $\alpha$), which can again be fitted to semi-inclusive pp cross sections. Following this logic, we can extend the GG model to enable us to distinguish between the two types of wounded nucleons, as the conditional probability for a wounded nucleon to also be diffractively excited, amounts to:
\begin{equation}
	\Theta\left(\sqrt{\sigma_{GG}/\pi} - (r_1 - r_2) - b\right) \frac{2 - \alpha}{2 - \alpha c}.
\end{equation}
\begin{figure}
	\centering
	\includegraphics[width=0.4\textwidth]{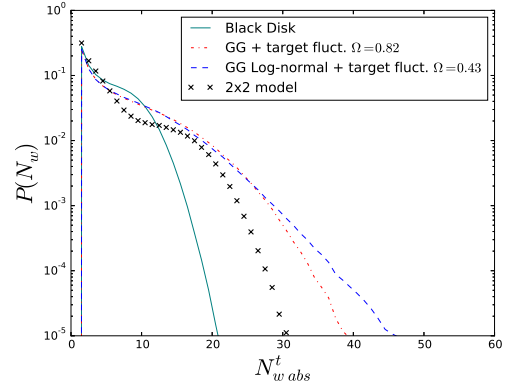}
	\caption{\label{fig:npartabs}Number of absorptively wounded participants in collisions of pPb at 5.02 TeV using four different Glauber models.}
\end{figure}
In figure \ref{fig:npart} we show the distribution of wounded nucleons for the GG model with added target fluctuations as per the above recipe for both the regular and the log-normal parametrization of $P_{tot}(\sigma)$. For comparison we show also the pure 2-radius model (named $2\times 2$) as well as ordinary black disk Glauber.
In figure \ref{fig:npartabs} we compare to the number of absorptively wounded participants. We see that the model retains their ordering as expected. We also see that adding fluctuations clearly produces a higher tail, both comparing fluctuating models to the black disk, but also comparing the log-normal distribution to the standard choice.
%\begin{figure}
%	\centering
%	\includegraphics[width=0.4\textwidth]{figures/pomeron.png}
%	\caption{\label{fig:pomeron}Pomeron diagrams with cuts indicated for (a) single diffractive excitation in proton--proton and (b) doubly absorptive proton--deuteron scattering.}
%\end{figure}
\begin{figure}
	\centering
	\includegraphics[width=0.4\textwidth]{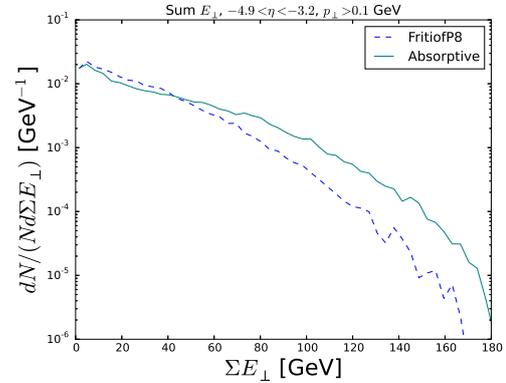}
	\caption{\label{fig:sumet}Distribution of $\sum E_\perp$ in the lead-going direction. used as a centrality observable by ATLAS.}
\end{figure}
\FloatBarrier
\section{Final states}
\subsection{Particle production}
We now want to estimate final states, by coupling the wounded nucleons to a model for particle production. A situation with just a single absorptively wounded nucleons is simple. It can simply be taken equal to a pp inelastic non-diffractive event, for which we will use the \pytppp MPI handling \cite{Sjostrand:1987su}. Adding another absorptive sub-collision does, however, not contribute equally much. The situation here is similar to doubly absorptive proton--deuteron scattering. \footnote{This is particularly visible if one considers the cut Pomeron diagrams of the two processes.} 

In the following we therefore treat additional absorptive sub-collisions \emph{as if} they were diffractive excitations. The real diffractive excitations are treated in the same way. High $p_\perp$ secondary absorptive exchanges are treated in a perturbative framework, whereas low $p_\perp$ are treated using a framework drawing from the old Fritiof event generator \cite{Andersson:1986gw}. This particle production model is dubbed "FritiofP8". For comparison we also show events where everything is treated as an absorptive sub-collision, dubbed "Absorptive".
\begin{figure}
	\centering
	\includegraphics[width=0.4\textwidth]{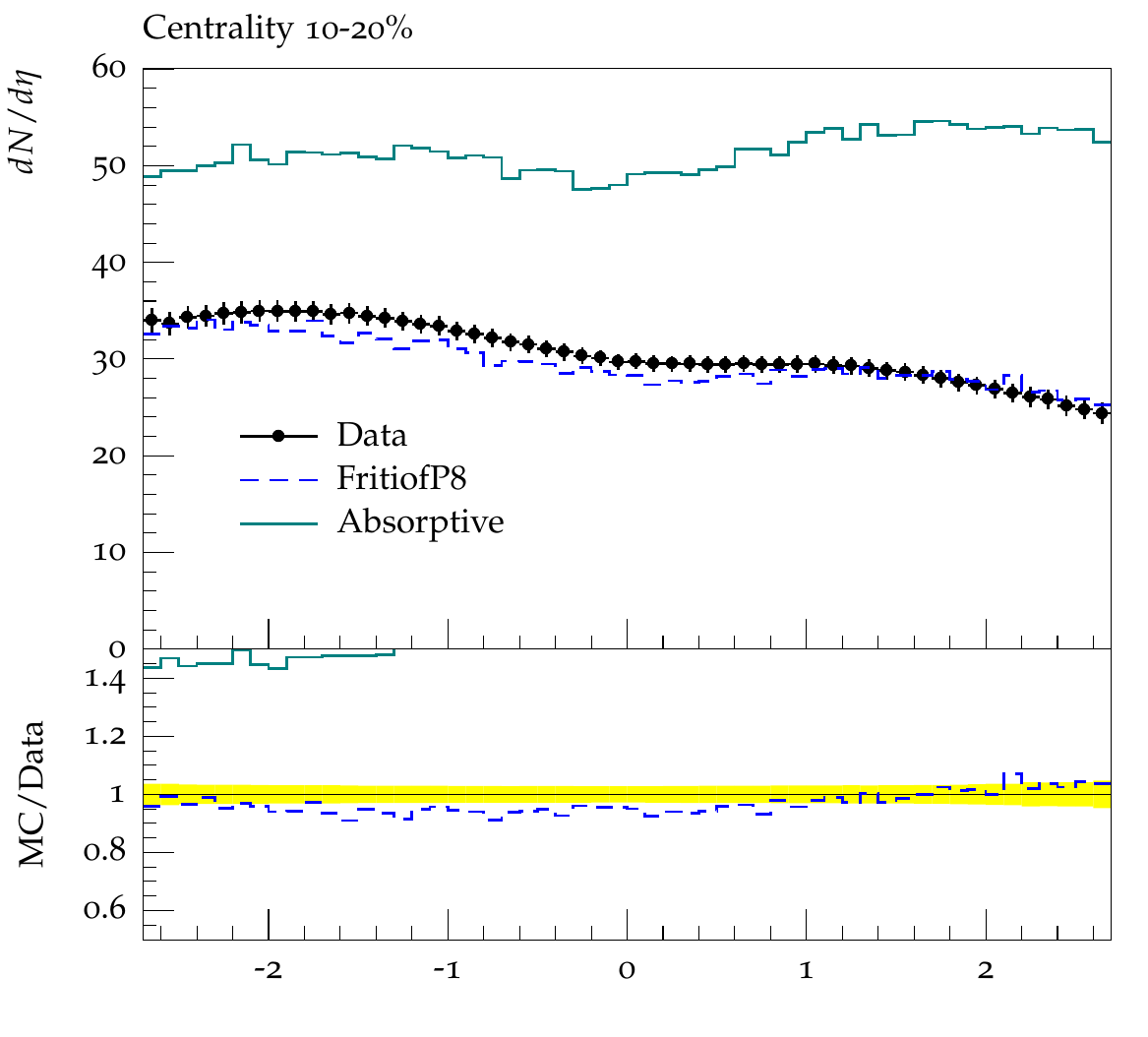}
	\caption{\label{fig:etaMiddle}Multiplicity distribution in $\eta$ for pPb, mid-centrality.}
\end{figure}
\subsection{Comparison to data}
Recent data by ATLAS \cite{Aad:2015zza} suggests the use of $\sum E_\perp$ in the forward direction as centrality observable. In figure \ref{fig:sumet} we show results for this distribution for both FritiofP8 and Absorptive. 
\begin{figure}
	\centering
	\includegraphics[width=0.4\textwidth]{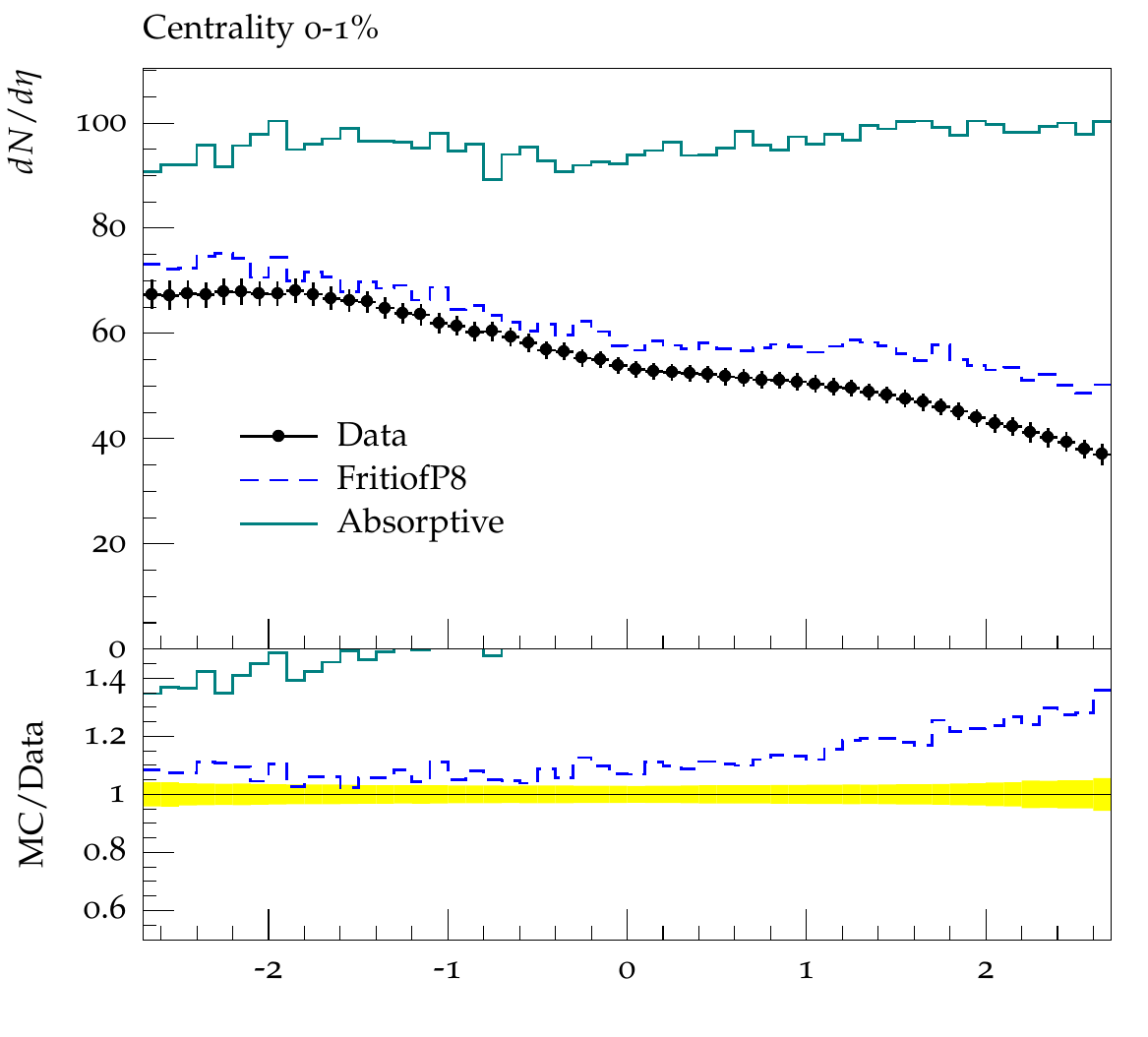}
	\caption{\label{fig:etaCent}Multiplicity distribution in $\eta$ for pPb, central events.}
\end{figure}
We see that the FritiofP8 model gives a softer distribution which is much more similar to data than the Absorptive. We bin in centrality by taking fractiles of this distribution, which gives multiplicity distributions in $\eta$, as shown in figure \ref{fig:etaMiddle} for mid-centrality events and figure \ref{fig:etaCent} for central events. 
We see that while the "absorptive" model overshoots, FritiofP8 manages to describe well the multiplicity in absolute numbers, the centrality dependance and the $\eta$-asymmetry.

\section{Acknowledgements}



%% References

\nocite{*}
\bibliographystyle{elsarticle-num}
\bibliography{refs}






% \end{thebibliography}

\end{document}